\documentclass[aps,prx,twocolumn,superscriptaddress,notitlepage]{revtex4-1}
\usepackage{graphicx}
\usepackage[caption=false]{subfig}
\usepackage{float}
\usepackage{natbib}
\usepackage{xcolor}
\usepackage{xr}
\usepackage{amsmath}
\usepackage{amssymb}
\usepackage{bm}
\usepackage{array}
\usepackage{wasysym}
\usepackage{color,soul}
\usepackage{braket}
\usepackage{verbatim}
\usepackage{hyperref}
\usepackage{gensymb}
\usepackage{url}
\usepackage{dirtytalk}
\usepackage{filecontents}
\usepackage{gensymb}

\makeatletter
\g@addto@macro\bfseries{\boldmath}
\makeatother
\def\cns{Co$_{1/4}$NbSe$_2$}
\def\musr{\texorpdfstring{$\mu^+$SR}\ }
\newcommand{\vect}[1]{\ensuremath{\bm{#1}}}
\newcommand{\unit}[1]{\ensuremath{\bm{\hat{#1}}}}
\newcommand*\diff{\mathop{}\!\mathrm{d}}

\begin{document}


\title{Local probe evidence supporting altermagnetism in \cns}

\author{J. N. Graham}
\email{jennifer.graham@psi.ch}
\thanks{These authors contributed equally to the work.}
\affiliation{PSI Center for Neutron and Muon Sciences CNM, 5232 Villigen PSI, Switzerland}

\author{T. J. Hicken}
\email{thomas.hicken@psi.ch}
\thanks{These authors contributed equally to the work.}
\affiliation{PSI Center for Neutron and Muon Sciences CNM, 5232 Villigen PSI, Switzerland}

\author{R. B. Regmi}
\affiliation{Department of Physics and Astronomy, University of Notre Dame, Notre Dame, IN 46556, USA}
\affiliation{Stravropoulos Center for Complex Quantum Matter, University of Notre Dame, Notre Dame, IN 46556, USA}

\author{M. Janoschek}
\affiliation{PSI Center for Neutron and Muon Sciences CNM, 5232 Villigen PSI, Switzerland}
\affiliation{Physik-Institut, Universität Zurich, Winterthurerstrasse 190, CH-8057 Zurich, Switzerland}

\author{I. Mazin}
\affiliation{Department of Physics and Astronomy, George Mason University, Fairfax, VA 22030, USA}
\affiliation{Quantum Science and Engineering Center, George Mason University, Fairfax, VA 22030, USA}

\author{H. Luetkens}
\affiliation{PSI Center for Neutron and Muon Sciences CNM, 5232 Villigen PSI, Switzerland}

\author{N. J. Ghimire}
\affiliation{Department of Physics and Astronomy, University of Notre Dame, Notre Dame, IN 46556, USA}
\affiliation{Stravropoulos Center for Complex Quantum Matter, University of Notre Dame, Notre Dame, IN 46556, USA}

\author{Z. Guguchia}
\email{zurab.guguchia@psi.ch}
\affiliation{PSI Center for Neutron and Muon Sciences CNM, 5232 Villigen PSI, Switzerland}


\begin{abstract}
Muon spin rotation (\musr), combined with muon stopping site and local field analysis, was used to investigate the magnetic properties of cobalt intercalated 2H-NbSe$_2$ (\cns). 
\cns\ is predicted to be an altermagnet, and whilst neutron diffraction has proposed its magnetic structure, microscopic details such as the magnetic volume fraction remain unclear. Therefore, a local probe investigation of its magnetism is essential.
Here, we report the determination of the magnetically ordered volume fraction, ordered moment size, and magnetic structure.
Our results reveal a sharp second-order transition to a full-volume-fraction, homogeneous magnetic order below $T_\mathrm{N}~=~168~$K.
The moments are aligned antiparallel along the $c$-axis, consistent with neutron diffraction and altermagnetism.
\musr\ reveals that the state remains stable under a \textit{c}-axis magnetic field up to $0.78~$T, with magnetisation measurements suggesting this robust regimes extends to at least $5$~T.
Within the time resolution of \musr, no precursor slow altermagnetic fluctuations were detected above $T_\mathrm{N}$, which is important for interpreting the band splitting in the paramagnetic state reported by photoemission studies.
These findings support altermagnetism in \cns\ and motivate further experiments to explore the tunability of its magnetic and electronic structure.
\end{abstract}
\maketitle

\section{Introduction}
Altermagnetism is a recently identified magnetic state that bridges the gap between conventional ferromagnetism and antiferromagnetism~\cite{PhysRevX.12.031042,PhysRevX.12.040501,PhysRevX.12.040002,doi:10.1073/pnas.2108924118,dale2024nonrelativisticspinsplittingfermi,regmi2024altermagnetismlayeredintercalatedtransition, wei2024crystal, doi:10.1021/jacs.4c14503}.
An altermagnet features unique spin-splitting behaviour, where time-reversal and crystal symmetries lead to momentum-dependent spin polarisation without net magnetisation.
A magnetic structure consisting of two antiparallel spin sublattices is also consistent with antiferromagnetism, however, the imposition of specific crystal symmetries leads to altermagnetism.
Altermagnets are characterised by intriguing properties such as a spin-polarised band structure, high mobility of charge carriers, spin-polarised transport without an external magnetic field, and robustness to external magnetic fields.
As such, altermagnets hold promise for spintronics applications, offering advantages like efficient spin current generation and manipulation. 

Several materials have been proposed as strong candidates for altermagnetism, where the specific crystal symmetries enable momentum-dependent spin splitting without a net magnetisation.
Currently, there is a pressing need to experimentally confirm altermagnetism in real materials.
Among the possible candidates are transition-metal compounds like CrSb~\cite{ding2024large, yang2025three} and MnTe~\cite{PhysRevB.107.L100418,jost2025chiralaltermagnonmnte}, which exhibit altermagnetic band structures.
Another promising platform for altermagnetism is the Lieb lattice~\cite{dürrnagel2024altermagneticphasetransitionlieb}, a bipartite structure characterised by a flat band and inherent sublattice asymmetry.
The Lieb lattice can host unconventional spin polarisation and strong electronic correlations, providing a fertile ground for realising altermagnetic states.

\begin{figure*}
    \centering
    \includegraphics[width=\textwidth]{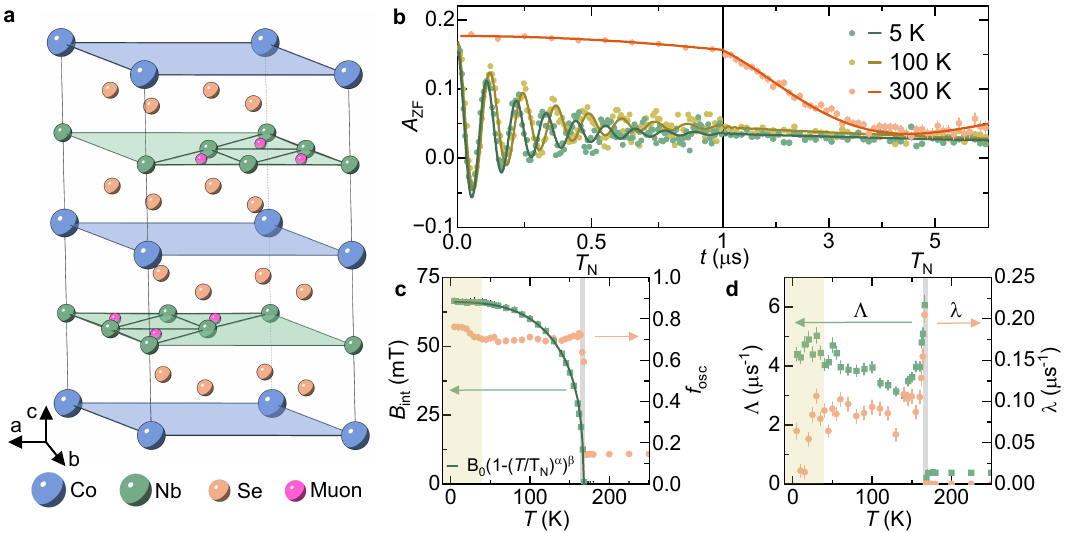}
    \caption{\textbf{Crystal structure, muon stopping site and summary of \musr\ experiments on \cns}. \textbf{a} \cns\ adopts the $P6_3/mmc$ crystal structure which consists of cobalt atoms (blue) being intercalated between layers of NbSe$_2$ (green and peach, respectively). The muon stopping site (and equivalent positions, pink) lies within the NbSe$_2$ plane (green) at the centre of the Nb triangle (lines are shown for clarity).  \textbf{b} Raw spectra of zero-field $\mu$SR data at $5$, $100$ and $300$~K. At low temperatures and short times there is a fast relaxation and oscillations indicating long-range magnetic order. \textbf{c} Internal field, $\mu_0H_\mathrm{int}$ (green) and magnetic volume fraction, $f_\mathrm{osc}$ (peach) as a function of temperature. \textbf{d} Evolution of static, $\Lambda$ (green) and dynamic, $\lambda$ (peach) relaxation rates with temperature. Yellow shaded region in \textbf{c} and \textbf{d} highlights the slight change in parameters below $30$~K, where the magnetisation exhibits a slight kink~\cite{regmi2024altermagnetismlayeredintercalatedtransition}.}
    \label{fig1}
\end{figure*}

Another class of materials that may host altermagnetism are the intercalated transition metal dichalcogenides (TMDs).
TMDs encompass a wide variety of $MX_2$ materials~\cite{wilson1969transition,wang2012electronics,manzeli20172d,choi2017recent,han2018van} where a transition metal layer ($M$) is sandwiched between two chalcogen layers ($X =$ S, Se or Te)~\cite{marseglia1983transition,manzeli20172d}.
One of the most studied TMD compositions is NbSe$_2$ which in bulk form ($2H$ trigonal prismatic stacking) exhibits a superconducting ground state, with critical temperature $T_\mathrm{C} = 7.2~$K, and develops a charge density wave (CDW) below $T_\mathrm{CDW} = 33.5~$K~\cite{wilson1975charge, soto2007electric}.
Both transitions can be greatly affected on approaching monolayer thickness, with the CDW increasing to an extraordinary $145~$K as $T_\mathrm{C}$ is suppressed to $3~$K~\cite{xi2015strongly}.
2H-NbSe$_2$, and related materials, are highly tunable under strain~\cite{wieteska2019uniaxialstraintuningsuperconductivity}, pressure~\cite{doi:10.1126/sciadv.aav8465,guguchia2017signatures}, or external fields, making them especially attractive for spintronics and to study exotic and emergent quantum states.

Intercalation offers further control over the electronic and magnetic properties, allowing fine-tuning of band structures, spin textures, and exchange interactions~\cite{jung2016intercalation,wang2020intercalated}.
In layered TMDs, magnetic intercalants can induce complex spin textures through spin-orbit coupling, electron correlation, and symmetry effects~\cite{moriya1982evidence,suzuki1993conduction,kousaka2016long,hall2021magnetic,zhang2021chiral,hicken2022energy,xie2022structure,eder2025structural}.
For example, two distinct magnetic phases can be produced by the cobalt intercalation of different concentrations~\cite{mandujano2025coexistence}.
Namely, when the concentration of the intercalant is $\frac{1}{3}$, (Co$_{1/3}$NbSe$_2$), the structure adopts the non-centrosymmetric $P6_322$ structure which subsequently induces an incommensurate magnetic structure below $T_\mathrm{N}~=~28~$K~\cite{mandujano2025coexistence}.
Alternatively, when the intercalant concentration is $\frac{1}{4}$ (e.g., \cns), the structure adopts the centrosymmetric $P6_3/mmc$ structure with moments believed to align antiparallel along the $c$-axis (two oppositely polarised ferromagnetic layers, akin to $A$-type antiferromagnetism) below $T_\mathrm{N}~=~168~$K~\cite{regmi2024altermagnetismlayeredintercalatedtransition}.

\cns\ has recently garnered attention as a strong candidate for altermagnetism~\cite{doi:10.1073/pnas.2108924118,PhysRevX.12.031042,PhysRevX.12.040002,PhysRevX.12.040501,regmi2024altermagnetismlayeredintercalatedtransition,dale2024nonrelativisticspinsplittingfermi}.
Density Functional Theory (DFT) calculations predict \cns\ to be an altermagnet~\cite{regmi2024altermagnetismlayeredintercalatedtransition}.
Experimental studies utilising a combination of magnetometry and neutron scattering show that \cns\ adopts an \textit{A}-type magnetic structure, with moments aligned antiparallel along the c-axis~\cite{regmi2024altermagnetismlayeredintercalatedtransition}.
This magnetic structure, with the wave vector \textbf{q}=0, coupled with the lack of inversion centre between two Co atoms due to the intermediary NbSe$_2$ layers, fulfils the requirements for an altermagnetic state in \cns\ below $168~$K.
Moreover, a recent study reports spin-ARPES~\cite{dale2024nonrelativisticspinsplittingfermi} measurements showing that spin splitting  mostly vanishes at $T_\mathrm{N}$, but some residual spin splitting extends into the paramagnetic state which is attributed to altermagnetic fluctuations. 
Despite the already reported magnetic properties, the microscopic details of the magnetic response, such as the magnetically ordered volume fraction, homogeneity of the magnetism, slow fluctuations both below and above $T_\mathrm{N}$, and the stability of the magnetic structure has not yet been settled. Addressing these aspects is of paramount importance. In this paper, we tackle these questions using a highly sensitive local probe technique muon spin rotation (\musr) combined with muon stopping site calculations and local field analysis. Our findings reveal a sharp second-order phase transition to a full-volume-fraction, homogeneous order below $T_\mathrm{N}$~=~168~K, with moments aligned antiparallel along the $c$-axis, indicating altermagnetism in this material.
Furthermore, the altermagnetic state remains stable under a $c$-axis magnetic field of $0.78~$T; considering this in the context of the linear reponse of magnetisation to an applied field~\cite{regmi2024altermagnetismlayeredintercalatedtransition}, this suggests the altermagnetic state is robust up to at least $5$~T.
This contrasts with transport measurements~\cite{devita2025opticalswitchinglayeredaltermagnet}, which indicate the onset of magnetic order below 50 K.
However, our results align with previous single-crystal neutron diffraction experiments and DFT predictions, providing strong support for robust altermagnetism in \cns~ below $T_\mathrm{N}$~=~168~K. Furthermore, within the time resolution of \musr, we find no evidence of precursor slow altermagnetic fluctuations above $T_\mathrm{N}$. This suggests that the persistent spin splitting above the ordering temperature does not originate from slow fluctuations at the MHz scale.

\section{Results}
Zero-field (ZF) \musr\  experiments were performed using the GPS instrument \cite{amato2017new} on a $c$-axis aligned mosaic arrangement of single-crystal samples of \cns. 
At high-temperatures, the ZF \musr\ spectra are well described by a Gaussian Kubo-Toyade function with a constant background, as expected for a system with no long-range magnetic order.
The background was estimated to be $17.0(3)~\%$ which can be attributed to any muons not stopping in the sample.
(Note that this is higher than typical for the GPS instrument; as the sample was relatively thin some muons passed straight through).
To confirm the magnetic volume fraction, we also performed weak-transverse-field measurements, where a field is applied perpendicular to the initial muon spin direction.
We find that the muons attributed to stopping in the sample all stop in regions of long-range magnetic order, suggesting that, within experimental uncertainty, the full-volume-fraction is magnetically ordered below $T_\mathrm{N}~=~168$~K.
Thus, we rule out the possibility of macroscopic phase separation into regions of different levels of Co intercalation, as has been suggested in Co$_{0.28}$NbSe$_2$~\cite{mandujano2025coexistence}.

Upon cooling, the ZF measurements show well resolved oscillations, characteristic of long-range magnetic order, for example, Fig.~\ref{fig1}b.
In principle, these data should be fitted with three parts: two arising from muons stopping in the sample (corresponding to muon-spin components perpendicular and parallel to the internal field), and one arising from muons stopping outside the sample.
In practice, the part arising from the component of the muon spin parallel to the internal field is indistinguishable from the background contribution, and hence the data were fitted with a Lorentzian damped internal field function
\begin{multline}
     A_\mathrm{ZF}^{\mathrm{GKT}}(t) = A\big[f\mathrm{_{osc}cos}\left(\gamma_\mu B_\mathrm{int}t \right)\mathrm{exp}\left(-\Lambda t\right) \\
     + \left(1-f_\mathrm{osc}\right)\mathrm{exp}\left(-\lambda t\right)\big] ,
\end{multline}
where $A$ is the initial asymmetry, $f_\mathrm{osc}$ is the oscillating fraction that arises from the component of muon-spins inside the sample that are perpendicular to the internal magnetic field $B_\mathrm{int}$, and $\gamma_\mu/(2\pi)~=~135.5$~MHz/T is the gyromagnetic ratio of the muon.
The depolarisation rates $\Lambda$ and $\lambda$ characterise the damping of the oscillating and non-oscillating part of the $\mu$SR signal respectively.
The damping $\Lambda$ arises due to a distribution of magnetic fields at the muon site, predominantly arising from static disorder.
The longitudinal relaxation rate $\lambda$ reflects depolarisation due to dynamic fluctuations.
The temperature evolution of the fit parameters are shown in Figs.~\ref{fig1}c and d, and are typical for the onset of a second order phase transition below $T_\mathrm{N}~=~168$~K.
Figure~\ref{fig1}c shows the gradual increase of the internal magnetic field below $T_\mathrm{N}$ which saturates at $67~$mT.
In contrast, the oscillating component of the phase fraction sharply increases below $T_\mathrm{N}$, and remains constant until a slight upturn below $30~$K (yellow shaded region), where the magnetisation also exhibits a small kink~\cite{regmi2024altermagnetismlayeredintercalatedtransition}. This has previously been noted as the temperature where charge order in \cns\ occurs~\cite{wieteska2019uniaxialstraintuningsuperconductivity}, or long-range magnetic order onsets for the Co$_{1/3}$NbSe$_2$ analogue \cite{mandujano2025coexistence}. 

\begin{figure}
    \centering
    \includegraphics[width=0.5\textwidth]{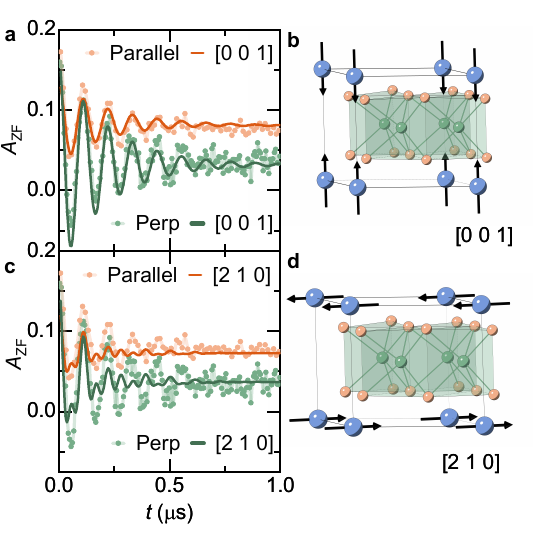}
    \caption{\textbf{Simulations of zero-field \musr\ spectra for different magnetic structures}. \textbf{a} and \textbf{b} Simulations against ZF \musr\ spectra and schematic of visual representation of structure for the accepted magnetic structure of \cns, moments aligned antiparallel along the $c$-axis. \textbf{c} and \textbf{d} Simulations against ZF \musr\ spectra and schematic of visual representation of structure for a moments aligned in the $[2~1~0]$ plane In \textbf{b} and \textbf{d} atoms are shown as Co = blue, Nb = green, Se = peach.}
    \label{fig2}
\end{figure}

We describe the internal field with a power law dependence
\begin{equation}
    B = B_0 \left(1-(T/T_\mathrm{N})^\alpha\right)^\beta
\end{equation}
where $B_0$ is the internal field at zero-temperature, $\alpha$ describes the scaling of low energy excitations, and $\beta$ is the critical exponent.
We obtain $\alpha~=~2.81(7)$ and $\beta~=~0.351(8)$, with $B_0~=~66.0(2)~$mT and $T_\mathrm{N}~=~167.0(2)$~K.
Whilst the error on $\beta$ makes it consistent with multiple spin types ($\beta~=~0.365$ for 3D Heisenberg spins, and $\beta~=~0.346$ for 3D $XY$-type spins), $\alpha$ is more informative.
One expects $\alpha~=~1.5$ for a ferromagnet and $2$ for an antiferromagnet~\cite{yaouanc2011muon}, hence our $\alpha>2$ suggests that \cns\ is consistent with a magnetic order where moments are aligned antiparallel, plus gapped spin wave excitations~\cite{blundell2022muon}.

The evolution of the muon spin relaxation rates $\Lambda$ and $\lambda$ are shown in Fig.~\ref{fig1}d.
As can be seen, the dominant relaxation is due to static disorder, with $\Lambda$ around forty times greater than the dynamic $\lambda$.
Below $30$~K, there is a decrease in $\lambda$, with an increase in the amplitude of the oscillating (and hence static) signal despite no change in the overall magnetic volume fraction.
As we are in the slow fluctuation limit, $\lambda~\propto~\nu$, where $\nu$ is the rate of magnetic fluctuations.
This indicates that at low $T$ we observe a slowing down of spin fluctuations, and a more static magnetic response.
We also observe that above $T_\mathrm{N}$ that the $\lambda$ rate falls to zero, which suggests the absence of precursor altermagnetic fluctuations on the \musr\ timescale. Our results restrict any precursor fluctuations to be very roughly above $\simeq~10^{12}$~Hz, or slower than $\simeq~10^4$~Hz \cite{blundell2022muon}. This suggests that slow fluctuations do not cause the persistent band splitting observed in the paramagnetic state, as reported by photoemission studies~\cite{dale2024nonrelativisticspinsplittingfermi}. However, the role of fast fluctuations, to which \musr\ is not sensitive, cannot be excluded as a contributing factor to band splitting.
 
To further understand the \musr\ spectra, we have computed the location of the muon stopping site in \cns, see Supplemental Material for further details.
We find a single site, shown in Fig.~\ref{fig1}a, located in the Nb layers, at the centres of Nb triangles adjacent to the Co site.
This local symmetry matches the muon site in a sister material, Cr$_{1/3}$NbS$_2$~\cite{hicken2022energy}, therefore suggesting that the muon stopping site may be robust to changes in intercalation in this series.
This muon stopping site, exactly in the Nb layer, with some Nb triangles unoccupied, can be understood by considerations of chemical bonding between the muon and the neighbouring Nb (see SM for details).

We have simulated the dipole field (note that the exchange-bias hyperfine field is zero by symmetry at this site) at the muon sites for various different magnetic structures, and compared the results with experimental data that were collected under a ZF setup with the muon spin rotated 45$^\circ$ with respect to the muon momentum, making it sensitive to the magnetism in both the $ab$ plane (perpendicular, green) and along the $c$-axis (parallel, orange) (Fig. \ref{fig2}).
We find that a magnetic structure where moments are aligned antiparallel and along the $c$-axis, provides excellent agreement between our calculations and the experimental data, as shown in Fig.~\ref{fig2}a.
This magnetic configuration is consistent with altermagnetism.
The simulations suggest a magnetic moment of $1.610(4)~\mu_\text{B}/\text{Co}$, only slightly higher than the $\sim1.374(98)~\mu_\text{B}/\text{Co}$ reported in Ref.~\cite{regmi2024altermagnetismlayeredintercalatedtransition} from neutron scattering.
Understanding the moment in intercalated TMDs is an ongoing and challenging problem~\cite{hawkhead2023band}; it has been suggested that the slightly peculiar moment in \cns\ may be explained by considering the complex crystal field splitting arising due to the Co local symmetry~\cite{dale2024nonrelativisticspinsplittingfermi}.
Further, the simulations predict that around 20\% of the muons are stopping in regions where there is no long-range order (for example, in the sample holder), consistent with the fitting shown in Fig.~\ref{fig1}c.
No other magnetic structure, such as those where the magnetic moments are pointing in the $ab$-plane, is able to reproduce such a good match to the experimental data or theoretical predictions.
For completeness, we show one such structure in Fig.~\ref{fig2}b, where the moments are aligned along the $[210]$ direction, where the predicted moment is over $3~\mu_\text{B}/\text{Co}$.

\begin{figure}
    \centering
    \includegraphics[width=0.5\textwidth]{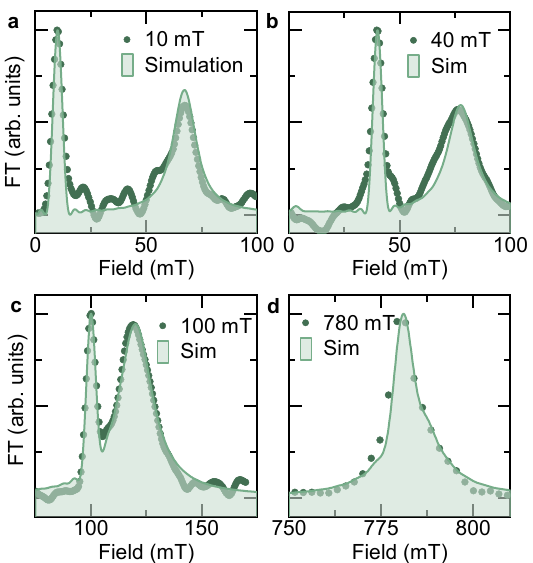}
     \caption{\textbf{Experimental and simulated TF-\musr\ data.} Comparison of experimental (dark green points) and simulated (light green shaded area) \musr\ data at $5~$K under applied transverse fields of \textbf{a} $10~$mT, \textbf{b} $40~$mT, \textbf{c} $100~$mT and \textbf{d} $780~$mT.}
    \label{fig3}
\end{figure}

Having understood the magnetic structure, we now turn to the effect of applying a magnetic field along the $c$-axis.
We have performed transverse-field (TF) \musr\ experiments where we are sensitive to the precession of the muon-spin in the vector sum of the internal and applied field.
Adding the relevant applied field to our calculated internal field for the $c$-axis aligned magnetic structure, we obtain predictions for the TF \musr\ spectra, which we compare to the experimental measurements in Fig.~\ref{fig3}.
Once again we obtain an excellent agreement by only slightly decreasing the background contribution at higher fields (which is expected experimentally due to beam focusing), and slightly changing the phenomenological damping factor.
This is further evidence that the magnetic structure is altermagnetic.
Further, we have demonstrated that no modification to the magnetic structure is needed to explain the \musr\ spectra up to 0.78~T.
The field dependence of the magnetisation in \cns~\cite{regmi2024altermagnetismlayeredintercalatedtransition} is linear between $0$ and $5$~T; thus, as \musr\ rules out any changes in the magnetic structure between 0 and 0.78~T, we conclude that the altermagnetic configuration is robust under $c$-axis-aligned applied fields of at least $5$~T.

Finally, we return to the low-temperature change in the \musr\ spectra. Having ruled out the possibility of macroscopic phase separation on the basis of our weak-transverse-field measurements, and additional dipole-field simulations (see SM), the origin of this feature must be intrinsic to \cns.
One explanation could involve nano-scale regions, possibly centred around extra Co inclusions, that remain dynamically fluctuating and disordered until lower temperatures.
As these regions undergo ordering, the average field at the muon site would remain largely unchanged, as observed in Fig.~\ref{fig1}c.
This transition would lead to a reduction in the dynamic response ($\lambda$ in Fig.~\ref{fig1}d), accompanied by an increase in the width of the field distribution at the muon site ($\Lambda$ in Fig.~\ref{fig1}d), due to the different symmetry of these Co inclusions. Alternatively, the characteristic temperature of 30~K coincides with the CDW transition temperature of NbSe$_{2}$.
If local structural distortions occur within the NbSe$_{2}$ layers below 30~K due to the onset of the CDW, these distortions could modify the bonding environment, thereby affecting the exchange coupling between Co inclusions.
Such an effect could lead to a local pinning of magnetic fluctuations by the CDW-induced lattice distortions, potentially influencing the observed magnetic dynamics.
This interplay between CDW order and local magnetic fluctuations suggests a possible coupling between the electronic and magnetic degrees of freedom, warranting further investigation through complementary techniques such as X-ray diffraction (XRD), and Raman spectroscopy to directly probe the structural modifications and their impact on magnetic order.

\section{Conclusion and outlook}
In this study, we employed the highly sensitive local magnetic technique \musr, combined with muon stopping site calculations, to elucidate the nature of the magnetic order of \cns.
Our findings reveal a sharp second-order phase transition to a $c$-axis, full-volume-fraction, homogeneous magnetic order below $T_\mathrm{N}=168$~K. The magnetic structure comprises moments aligned antiparallel along the 
$c$-axis, which is consistent with the previous neutron scattering experiment and the emergence of altermagnetism. Additionally, we find no evidence of precursor slow altermagnetic fluctuations above $T_\mathrm{N}~=~168$~K within the time-resolution of \musr. This indicates that the persistent spin splitting above the ordering temperature is not caused by slow fluctuations occurring at the MHz scale. Furthermore, we demonstrate that the width of the distribution of the fields at the muon site has a very weak dependence on the applied field.
This highlights the robustness of the altermagnetic state in \cns\ up to an applied field of $0.78$~T (and probably $5$~T) when applied along the $c$-axis, consistent with magnetisation experiments.
This robust behaviour is important both for the opportunity to employ other experimental probes that require large applied fields whilst still faithfully probing the altermagnetic state, and for potential spintronic applications. 

Having established a robust altermagnetic state in \cns, several important questions remain which are to be addressed in future studies.
A key manifestation of altermagnetism is its potential to exhibit an anomalous Hall effect (AHE).
In \cns\ the AHE is only expected when the easy axis lies within the \textit{ab} plane and is perpendicular to the Co–Co bond \cite{regmi2024altermagnetismlayeredintercalatedtransition}. As these conditions are not met, no AHE will be observed.
However, the strong experimental evidence supporting altermagnetism in this material motivates future investigations aimed at tuning the magnetic structure through uniaxial strain, hydrostatic pressure, high magnetic fields, and other external stimuli.
Such tuning could induce a transition from an out-of-plane to an in-plane configuration, thereby realising AHE.
If achieved, this would provide a comprehensive demonstration of altermagnetic properties, further solidifying the system's potential for spintronics applications.
Moreover, this tunability presents an opportunity to explore the unresolved question of phonon-magnon coupling—specifically, how atomic displacements influence exchange interactions, and conversely, how magnetic order affects lattice dynamics.

Beyond \cns\ there remains the opportunity to exercise further control of the magnetic properties of TMDs via tuning of the intercalant. 
As has been shown for Co-intercalated NbSe$_2$, slight changes in the stoichiometry of the Co can dramatically change the physical properties of the system \cite{mandujano2025coexistence}. 
This idea may be extended to include the vast number of combinations between TMDs and intercalcants of different varieties such as transition-metal ions, organic linkers and alkali metals \cite{wilson1969transition,wang2012electronics,manzeli20172d,choi2017recent,han2018van,jung2016intercalation,wang2020intercalated}. 
Addressing these fundamental aspects would provide deeper insights into the intrinsic interactions governing altermagnetic materials, potentially revealing novel couplings relevant to quantum materials and next-generation electronic devices.

\section*{Acknowledgments}
The ${\mu}$SR experiments were carried out at the Swiss Muon Source (S${\mu}$S) Paul Scherrer Insitute, Villigen, Switzerland. Z.G. acknowledges support from the Swiss National Science Foundation (SNSF) through SNSF Starting Grant (No. TMSGI2${\_}$211750).\\

\section*{Supplemental Material}
\subsection{Experimental details}
Muon spin rotation and relaxation (\musr) measurements of \cns\ were collected on the General Purpose Surface-Muon (GPS) instrument~\cite{amato2017new} at the Swiss Muon Source, Paul Scherrer Institute, Villigen, Switzerland.
Single-crystal samples were placed in a mosaic arrangement (i.e. the $c$-axis was aligned, without alignment in the $ab$-plane), and the muon spin was rotated to a $45^\circ$ so that measurements were sensitive to both the $ab$ plane and $c$-axis.
A continuous flow cryostat was used to control the temperature across a range of $5~$K to $300~$K.
Measurements were conducted in zero-field (ZF), and transverse-fields (TF) between $3~$mT and $0.78~$T.
Data were analysed using the \textsc{musrfit} software \cite{musrfit}.

The muon stopping sites in \cns\ were calculated using a DFT$+\mu$ approach~\cite{blundell2023dft} using the MuFinder application~\cite{huddart2022mufinder}.
Density functional theory calculations were performed using the plane-wave pseudopotential code \textsc{castep}~\cite{clark2005first}.
The PBEsol functional~\cite{perdew2008restoring} was used in all calculations.
Calculations were converged to $2.5$~meV/atom using a plane-wave cutoff of 600~eV and a $3\times3\times3$~$k$-point grid~\cite{monkhorst1976special}.
Initially, 36 random positions were initialised in the conventional unit cell ($\sim7\times7\times12$~\AA).
The atomic positions were subsequently allowed to relax until the energy and atomic positions have converged to better than $1\times10^{-3}$~meV/atom and $1\times10^{-4}$~\AA\ respectively, and the maximum force on any atom is less than $2\times10^{-3}$~eV/\AA.
The final muon positions are clustered using the MuFinder application to give the candidate muon stopping sites; each cluster was subsequently used to initialise a calculation in a $2\times2\times1$ supercell ($\sim14\times14\times12$~\AA) to ensure the muons did not self-interact.
The final muon stopping site was used in the \textsc{muesr} code~\cite{bonfa2018introduction} to calculate the dipolar field at these sites for candidate magnetic structures.
We have accounted for the sample orientation (a mosaic of crystals aligned along the $c$-axis, with no alignment in the $ab$-plane), see below for details, which means we can simultaneously predict the time-dependent spin-polarisation in two different detector pairs directly from the calculated vector field at the muon site.
This has the advantage of giving sensitivity to internal field directions, as well as magnitudes, which further helps compare different potential magnetic structures.
In our simulations, we assume a fraction of the muons stop outside of the sample, and apply a phenomenological damping factor to account for the distribution of fields at the muon site that likely arises due to the finite magnetic domain size. Finally, we allow the non-precessing component of the signal to weakly relax, accounting for any dynamics in the material.

\begin{table}[]
\begin{tabular}{c|c|c|c}
\textbf{Site} & \textbf{Energy (eV)} & \textbf{Coordinates} & \textbf{Wyckoff site} \\ \hline \hline
1 & 0 & (0.83, 0.66, 0.25) & $6h$\\
2 & 1.04 & (0.53, 0.06, 0.10) & $12k$\\
3 & 1.37 & (0.27, 0.13, 0.01) & $12k$\\
4 & 1.58 & (0.18, 0.09, 0.09) & $12k$\end{tabular}
\caption{\textbf{The calculated muon stopping sites in Co$_{1/4}$NbSe$_2$.} Energies are given with respect to the lowest energy site, which is beleived to be the only one occupied. Corrdiantes are fractional coordinates in the conventional unit cell.}
\label{tab:muonSites}
\end{table}

\subsection{Determination of the muon stopping site in \cns}
Our DFT calculations reveal four potential stopping sites, given in Table~\ref{tab:muonSites}.
The lowest energy site causes almost no distortion to the crystal structure, however the higher energy sites are slightly more disruptive and significantly higher in energy, suggesting they are less likely to be occupied.
Although the crucial parameter for determining occupation is the energy barrier between sites, rather than the energy of the local minima that are individual stopping sites, it would be surprising for sites 2--4, where the local minima is over 1~eV higher in energy, to be occupied.
Independent on the choice of magnetic structure, when simulating the field at the different muon sites, we find that the field at sites 3 and 4 (situated near Co atoms) are implausibly high, and thus can independently be assumed to be unoccupied.
The experimental ZF \musr\ spectra presented in the main text show the same, single frequency in both detector pairs, further suggesting only a single muon-stopping site.
Most significantly, our excellent agreement between the calculations and experimental data is achieved with 100\% occupation of the lowest energy muon site; we therefore conclude that this is the only occupied site in experiment.

\subsection{Derivation of the \musr\ response to a partially orientated mosaic of crystals}
In \musr\ experiments, one is often able to align a single crystal axis, with no knowledge of the alignment in the other directions.
As such, it is useful to explicitly derive the \musr\ response to such a partially orientated mosaic of single crystals.
We start with the general vector form of the time-dependent polarisation $\vect{P}(t)$ arising from a muon stopping in a single internal field $\vect{B}_\mu$~\cite{amato2024introduction}
\begin{multline}
    \vect{P}(t) = \left[\vect{P}(0)\cdot\unit{b}_\mu\right]\unit{b}_\mu \\
    + \left\{\unit{b}_\mu\times\left[\vect{P}(0)\times\unit{b}_\mu\right]\right\}\cos\left(\gamma_\mu B_\mu t\right) \\
    + \left[\vect{P}(0)\times\unit{b}_\mu\right]\sin\left(\gamma_\mu B_\mu t\right) ,
\end{multline}
where $\vect{P}(0)$ is the initial polarisation of the muon spin (assumed to be a unit vector), $\unit{b}_\mu=\vect{B}_\mu/\left|\vect{B}_\mu\right|=\vect{B}_\mu/B_\mu$, and $\gamma_\mu=2\pi\times135.538810(3)$~MHz/T is the gyromagnetic ratio of the muon.
Typically the initial spin polarisation is taken to be along $\unit{z}$, however we employ a different convention, where the aligned sample axis is along $\unit{z}$, with the initial spin polarisation set with respect to this.
Further, it is helpful to recast our problem in spherical polar coordinates, i.e. $\unit{b}_\mu~=~\left(\sin\theta\cos\phi, \sin\theta\sin\phi, \cos\theta\right)$.

From this equation, one is able to perform various averages.
The most familiar is the powder average, assuming a completely unoriented sample
\begin{equation}
    \vect{P}_\text{powder}(t) = \frac{1}{4\pi}\int_{\theta=0}^{\theta=\pi}\int_{\phi=0}^{\phi=2\pi}\vect{P}(t)\sin\theta\diff\phi\diff\theta .
\end{equation}
Taking $\vect{P}(0)=(1/\sqrt{2},0,1/\sqrt{2})$, which corresponds to the experimental setup used in this paper, specifically a spin-rotation of 45$\degree$ away from the incoming muon-momentum direction, we gain the expected result
\begin{align}
    \vect{P}_\text{powder}(t) &= \left(\begin{matrix}
    \frac{1+2\cos\left(\gamma_\mu B_\mu t\right)}{3\sqrt{2}} \\
    0 \\
    \frac{1+2\cos\left(\gamma_\mu B_\mu t\right)}{3\sqrt{2}}
    \end{matrix}\right) ,
\end{align}
i.e. the same signal in two perpendicular directions (corresponding the the forward-backward and up-down detector pairs) with an appropriate weighting, and no signal in the third direction.
Similarly, the relevant integral for the partially aligned mosaic is
\begin{equation}
    \vect{P}_\text{mosaic}(t) = \frac{1}{2\pi}\int_{\phi=0}^{\phi=2\pi}\vect{P}(t)\diff\phi ,
\end{equation}
which gives, using the same $\vect{P}(0)$,
\begin{align}\label{eqn:Pmosaic_beforeSymmetry}
    \vect{P}_\text{mosaic}(t) &= \left(\begin{matrix}
    \frac{\left(3+\cos2\theta\right)\cos\left(\gamma_\mu B_\mu t\right) + 2\sin^2\theta}{4\sqrt{2}} \\
    -\frac{\cos\theta\sin\left(\gamma_\mu B_\mu t\right)}{\sqrt{2}} \\
    \frac{\cos^2\theta + \sin^2\theta\cos\left(\gamma_\mu B_\mu t\right)}{\sqrt{2}}
    \end{matrix}\right) .
\end{align}
This gives the surprising result that there should be a time dependent polarisation in all three directions, which one will not observe experimentally.
To account for this, we must also consider alignment of the sample under the transformation $\unit{z}\mapsto-\unit{z}$, which produces the same result as Eqn.~\ref{eqn:Pmosaic_beforeSymmetry}, but with the second component multiplied by $-1$.
Averaging over these two results gives our final result
\begin{align}
    \vect{P}_\text{mosaic}(t) &= \left(\begin{matrix}
    \frac{\left(3+\cos2\theta\right)\cos\left(\gamma_\mu B_\mu t\right) + 2\sin^2\theta}{4\sqrt{2}} \\
    0 \\
    \frac{\cos^2\theta + \sin^2\theta\cos\left(\gamma_\mu B_\mu t\right)}{\sqrt{2}}
    \end{matrix}\right) .
\end{align}

\begin{figure}
    \centering
    \includegraphics[width=0.4\textwidth]{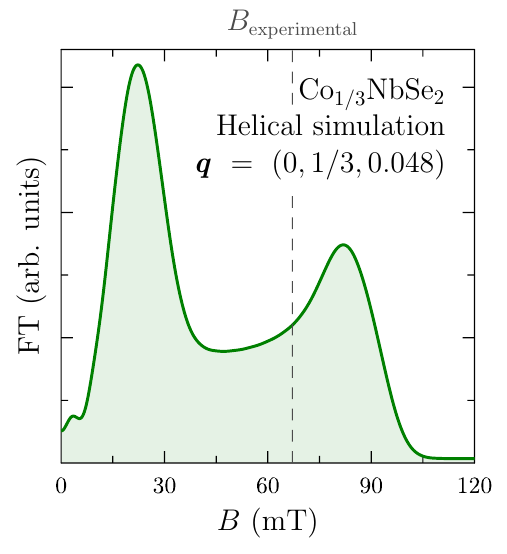}
     \caption{\textbf{Simulated field at the muon site in Co$_{1/3}$NbSe$_2$.} The field measured experimentally in this work at 5~K is also indicated.}
    \label{figS1}
\end{figure}

\subsection{Simulations of the field at the muon site in Co$_{1/3}$NbSe$_2$}
One suggestion~\cite{mandujano2025coexistence} that for the observed low-$T$ changes in the ZF \musr\ spectra is that a small fraction of the sample is actually closer to Co$_{1/3}$NbSe$_2$, leading to spatially separated magnetic domains at low-$T$ when these regions magnetically order.
To explore this possibility, we have calculated the field at the muon site in Co$_{1/3}$NbSe$_2$, assuming the proposed helical structure~\cite{mandujano2025coexistence}, which predicts a helix with $\vect{q}~=~(0, 1/3, 0.048)$, and assume the same magnetic moment as found in the main text, specifically $1.61~\mu_\text{B}/\text{Co}$.
We follow the same approach as set out for \cns, however as the structure is incommensurate, the fields at the muon site are calculated many times, with the initial phase of the magnetic structure varied continuously; this captures the effect of muons stopping in all different unit cells.
We have assumed the muon site is the same as that found for isostructural Cr$_{1/3}$NbS$_2$~\cite{hicken2022energy} (which shares the same local symmetry as the muon site in \cns), as is often the case for structurally similar compounds~\cite{franke2018magnetic}.
As is typical in materials with a helical magnetic structure, the internal field distribution seen by \musr\ comprises of two peaks (centred around 20 and 85 mT) with a continuous distribution between them, as shown in Fig.~\ref{figS1}.
As can be seen, the two peak fields are higher and lower than the experimentally measured internal field.
These features would be very obvious in the experimental data (even at a small sample volume), therefore, as we do not observe these features, we conclude that this is unlikely to be an explanation for the low-$T$ changes in the asymmetry spectra.

\subsection{Additional notes on the stopping sites in \cns}
In recent years, accurately calculating the muon stopping site~ has become routinely possible using the DFT+$\mu$\ method~\cite{blundell2023dft} (where the muon is treated as a reduced mass proton) using applications such as MuFinder~\cite{huddart2022mufinder}.
This method, whilst considerably more accurate than previous approaches, has made it harder to have good intuition about the stopping site.
In contrast, an older, often inaccurate method, where the muon was assumed to stop at the electrostatic potential minimum with no distortion to the surrounding lattice, was easier to predict `by eye'.
In this section we provide some additional insights into the muon stopping site in \cns\ which may provide improved intuition in other materials.

\begin{figure}
    \centering
    \includegraphics[width=0.4\linewidth]{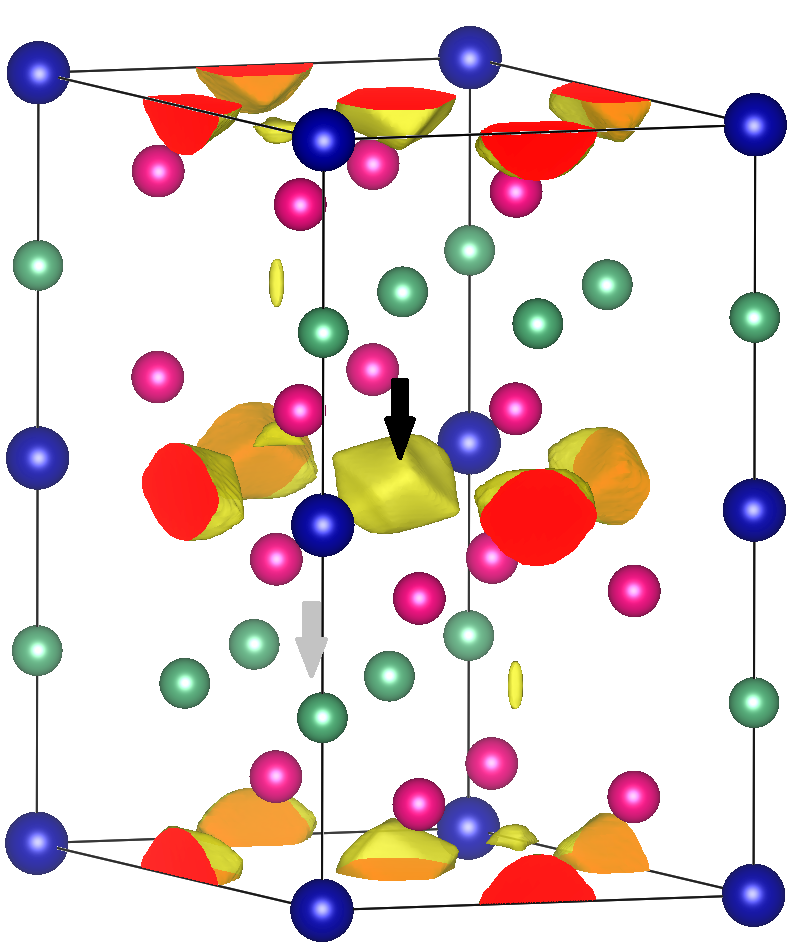}
    \includegraphics[width=0.55\linewidth]{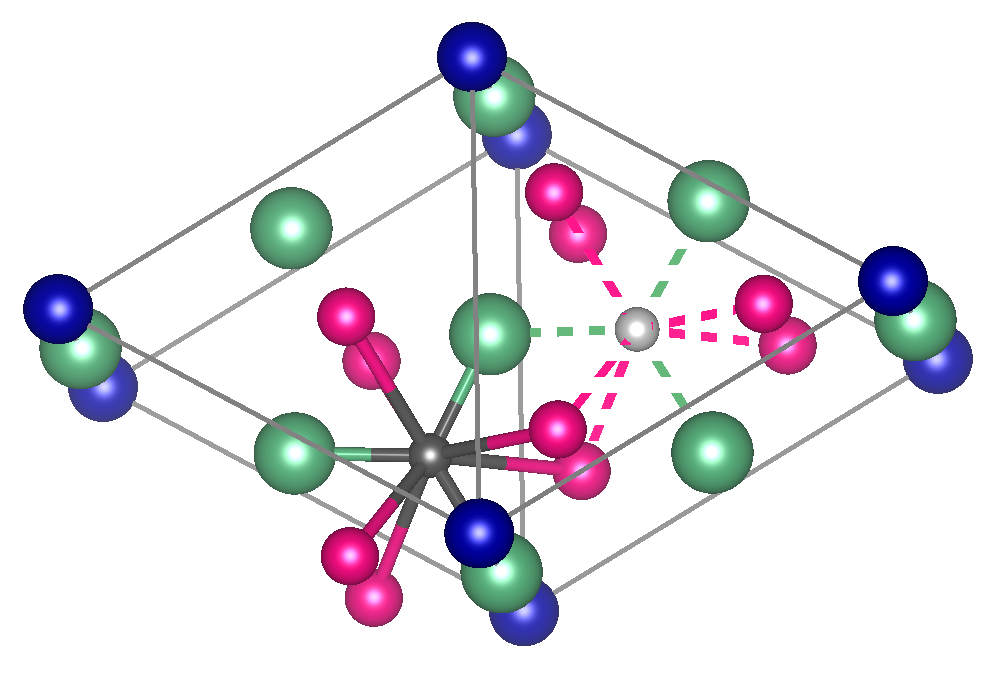}
    \caption{Left: Electrostatic potential. The black arrow shows the location of the minimal potential, whereas the grey arrow shows the actual calculated stopping site. Nb is shown in red, Se in green and Co in blue. Right: The calculated stopping site (black) vs. an alternative site (which has a higher energy) with the same $\mu^+$-Nb and  $\mu^+$-Se coordination.}
    \label{fig:electrostatic}
\end{figure}

If one calculates the electrostatic potential in \cns, one finds the minimum of the Coulomb potential occurs in the Co plane, due to the effect of the positive charge of Nb.
This is shown in Fig.~\ref{fig:electrostatic}.
An intuitive picture is therefore that the positively charged muon is attracted to this position, where it then tries to bond with the nearest ligand, stopping at a low-symmetry site.
Conversely, our muon site is actually a high symmetry position (as is often found in metals, for example Cu~\cite{luke1991muon}), in the Nb plane.
This site, which does not have the $z$ coordinate fixed by symmetry, is found to be exactly in the plane within computational accuracy; this means that the contact hyperfine coupling is zero by symmetry, and only the dipole contribution remains.
To gain an intuitive understanding, we therefore must pose two questions.
(1) Why does the muon prefer to stop inside the Nb plane, despite a higher electrostatic potential?
(2) Why of the two equivalent (apart from Co) positions in the Nb plane does the muon prefer the one that is closer to Co?

\begin{figure}
    \centering
    \includegraphics[width=0.4\linewidth]{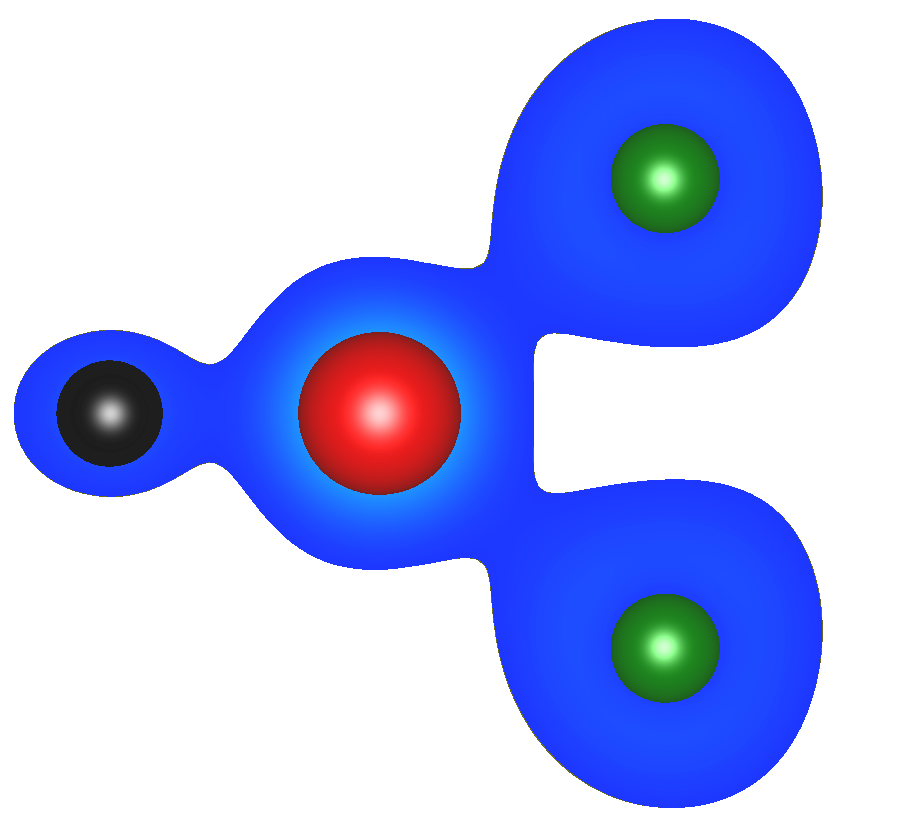}
    \includegraphics[width=0.55\linewidth]{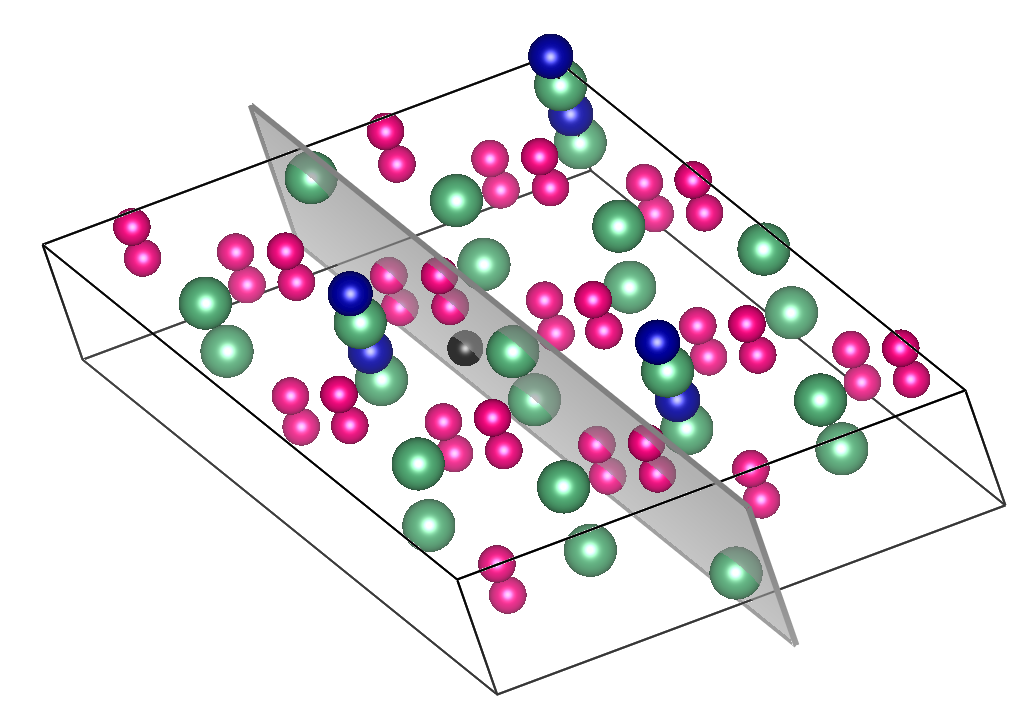}
    \caption{Charge density isosurfaces (at $n=0.05$ e/$\AA^3$). On the left is a cut which is shown in the full cell on the right. The muon is shown in black, with the Nb in red, Se in green and Co in blue.}
    \label{fig:charge}
\end{figure}

To answer question (1), we have calculated the charge distribution, as shown in Fig.~\ref{fig:charge}.
This charge distribution suggests that there is a chemical bond between the $\mu^+$ and the neighbouring Nb atoms which is of similar strength to the Nb-Se bond.
Thus, in this compound the location of the stopping site is primarily determined not by electrostatic energy, but by chemical bonding.

Now considering question (2), we show in Fig.~\ref{fig:electrostatic} two potential stopping sites inside the Nb plane.
The realised site is in a region of higher electrostatic potential when compared to the unoccupied position, albeit not by much.
One can again understand this by considering the chemical bonding between the $\mu^+$ and the surrounding Nb.
The length of the bonds that the $\mu^+$ forms with Nb (solid tubes in Fig.~\ref{fig:electrostatic}) are about 0.1~\AA   ~shorter than the dashed lines for the hypothetical alternative site; thus this bond is stronger, resulting in this site being occupied in preference to the alternative possibility.
Thus, the muon position in \cns\ can be intuitively understood by considering the chemistry of the system, with the electrostatic potential not playing an important role.

\bibliography{References}{}
\end{document}